\input{aipcheck}

\documentclass[
    ,final            
  ]
  {aipproc}
\layoutstyle{6x9}

\begin{document}

\title{UVES - VLT High Resolution Spectroscopy of GRB Afterglows}


\author{V. D'Elia}{
  address={INAF - Osservatorio Astronomico di Roma, Italy}
}

\author{S. Piranomonte}{
  address={INAF - Osservatorio Astronomico di Roma, Italy}
}

\author{P. Ward}{
  address={Dunsink Observatory, Castelknock, Dublin, Ireland}
}
\author{F. Fiore}{
  address={INAF - Osservatorio Astronomico di Roma, Italy}
}
\author{E.J.A. Meurs}{
  address={Dunsink Observatory, Castelknock, Dublin, Ireland}
}
\author{L. Norci}{
  address={School of Physical Sciences, Dublin City University,
Dublin, Ireland}
}
\author{S.D. Vergani}{
  address={Dunsink Observatory, Castelknock, Dublin, Ireland}, altaddress={School of Physical Sciences, Dublin City University,
Dublin, Ireland}
}

\begin{abstract}
We present early time, high resolution spectroscopy of three GRB
afterglows: GRB050730, GRB050922C and GRB060418.  These data give us
precious information on the kinematics, ionization and metallicity of
the interstellar matter of GRB host galaxies up to a redshift z $\sim
4$, and of intervening absorbers along the line of sight. 

In particular, we find that the GRB surrounding medium is complex,
featuring a clumpy environment in which gas is distributed in regions
of space with different distances from the GRB explosion site.  The
absorption spectra show that elements are present both with high and
low ionization states, and even forbidden, fine structure levels are
commonly observed. These features allow us to evaluate the physical
parameters of the absorbing gas. In details, the density of the gas
regions lie in the range $n = 10 \div 10^{6}$ cm$^{-3}$, and the
temperatures are of the order of $T = 10^3 \div 10^{4}$ K.

The metallicity of the GRB host galaxies is computed using the
hydrogen absorption features. We find undersolar abundances for our
GRBs, namely, $Z_{odot} \sim 10^{-3} \div 10^{-2}$. However,
$Z_{odot}$ can be underestimated since the H column presents large
uncertainties and dust depletion has not been taken into account. The
latter effect can be taken into account using as metallicity
indicators Zn and Cr, which tend to remain in the gas phase. We find
metallicities higher than the previous values and in agreement with
other measurements for GRB host galaxies.

Finally, the observed [C/Fe] ratio for GRB050730 ($z \sim
4$) agrees with values expected for a galaxy younger than a Gyr
undergoing bursts of star-formation. In addition, the [C/Fe] ratio
evaluated component by component can give informations on the relative
distances of the components from the GRB explosion site, since Fe dust
is more efficiently destroyed than graphite; inversely, if the
distance of the shells from the centre were known, we could obtain a
powerful tool to investigate the dust depletion in GRB host galaxies.

\end{abstract}

\maketitle


\section{Introduction}

Gamma Ray Bursts (GRBs) are one of the great wonders of Universe. They
combine several of the hottest topics of 21$^{st}$ century
astrophysics. On one side they are privileged laboratories for
fundamental physics, including relativistic physics, acceleration
processes and radiation mechanisms. On the other side, being GRB
associated to the death of massive stars, they can be used as a
cosmological tool to investigate star-formation and metal enrichment
at the epochs of galaxy birth, formation and growth.

Both purposes can be achieved studying the UV absorption lines of the
GRB afterglow spectra. In fact, early time spectroscopy of GRB
afterglows can give us precious information on the kinematics,
ionization and composition of the gas surrounding the GRB to put
constrains on the GRB progenitor models.

In addition, we can characterize the physical and chemical status of
the matter along the line of sight using the GRB as a background
beacon for spectroscopy of UV lines. up to a redshift $z
\sim 4$. Absorption line spectroscopy is certainly a powerful
tool. Spectroscopy of GRBs can be used to study
both the galaxy Inter-Stellar-Matter (ISM) and the
Inter-Galactic-Matter (IGM) along the line of sight. 

High resolution spectroscopy is important for many reasons:
1) absorption lines can be separated into several components
belonging to the same system; 2) the metal column densities can
be measured through a fit to the line profile for each component;
3) faint lines can be observed.

We focus here on the study of the local, GRB surrounding medium and
on the analysis of the absorption lines produced by the
gas in the host galaxy throughout observations of three GRBs:
GRB050730, GRB050922C and GRB060418.

\section{Observations}

The GRB050730 afterglow was observed 4.0 hours after the trigger. We find
seven main absorption systems at z=3.968, 3.564, 2.2536, 2.2526,
2.2618, 1.7729 and 1.7723

The GRB050922C afterglow was observed 3.5 hours after the trigger. We find
four main absorption systems at z=2.199, 2.077, 2.008 and 1.9985.

The GRB060418 afterglow was observed 10 minutes after the trigger. We
find four main absorption systems at z=1.489, 1.106, 0.655 and 0.602.

\begin{figure}
\vbox{

\includegraphics[height=15cm,width=6cm, angle=-90]{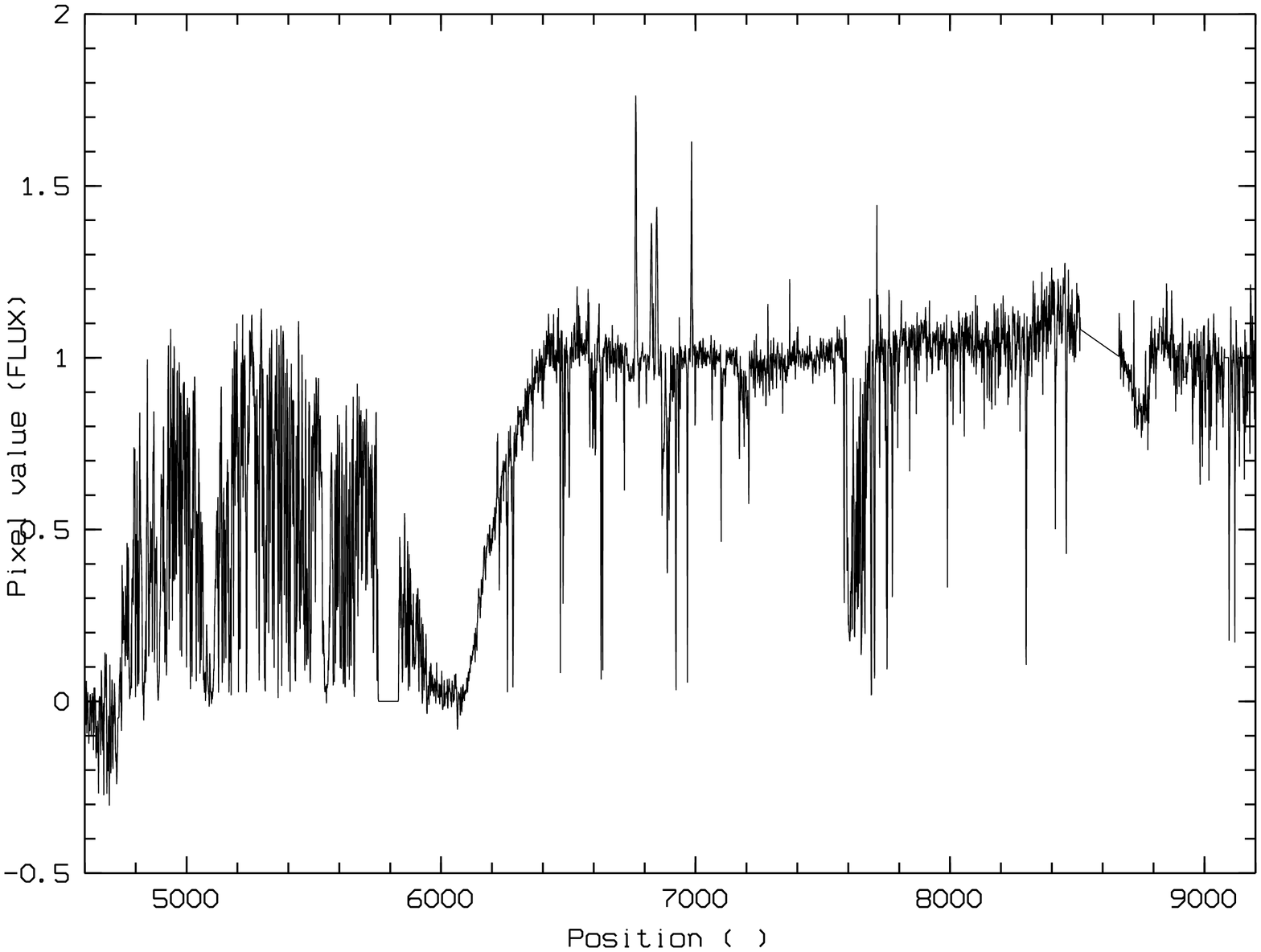}

\includegraphics[height=15cm,width=6cm, angle=-90]{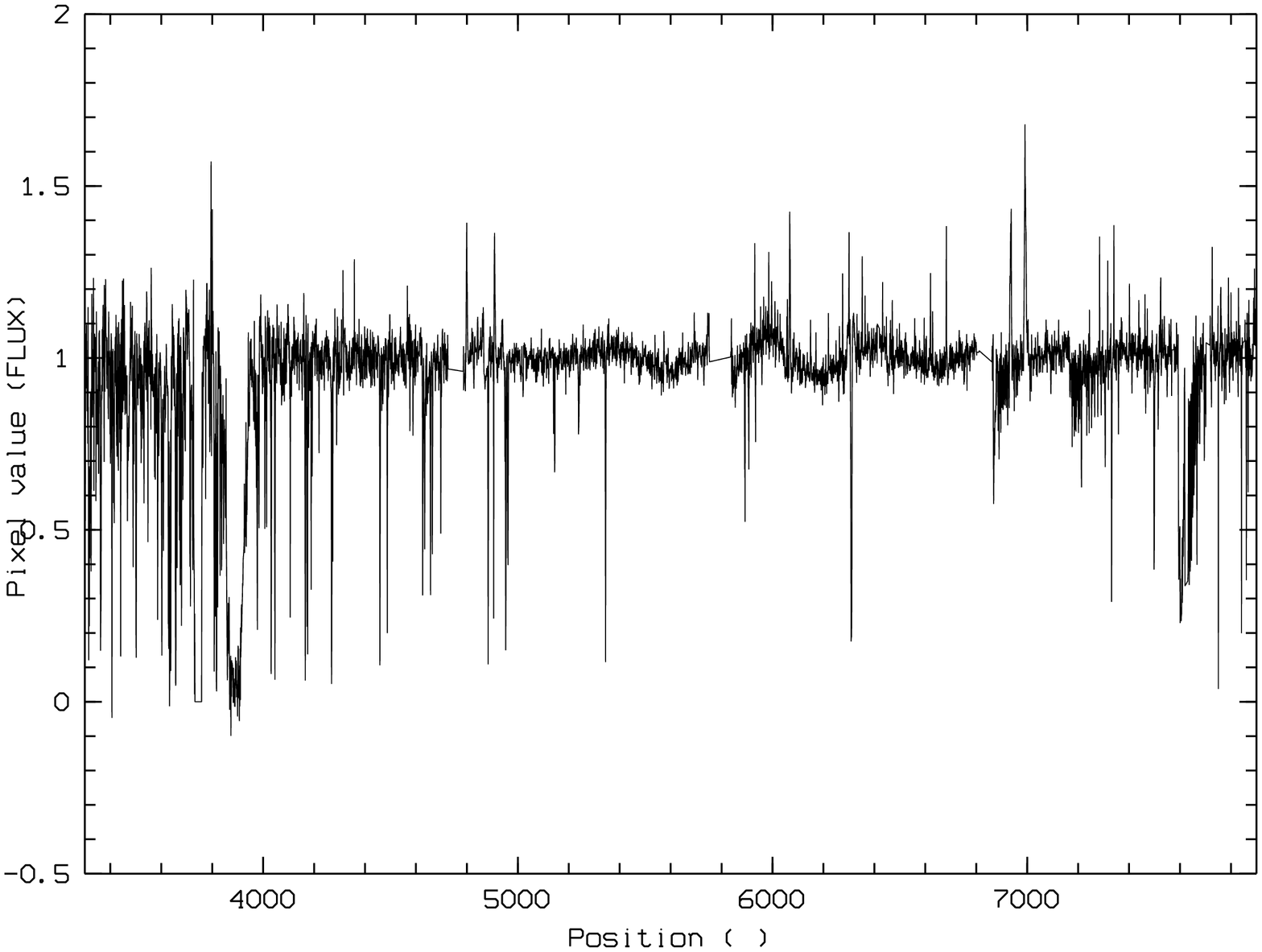} 

\includegraphics[height=15cm,width=6cm, angle=-90]{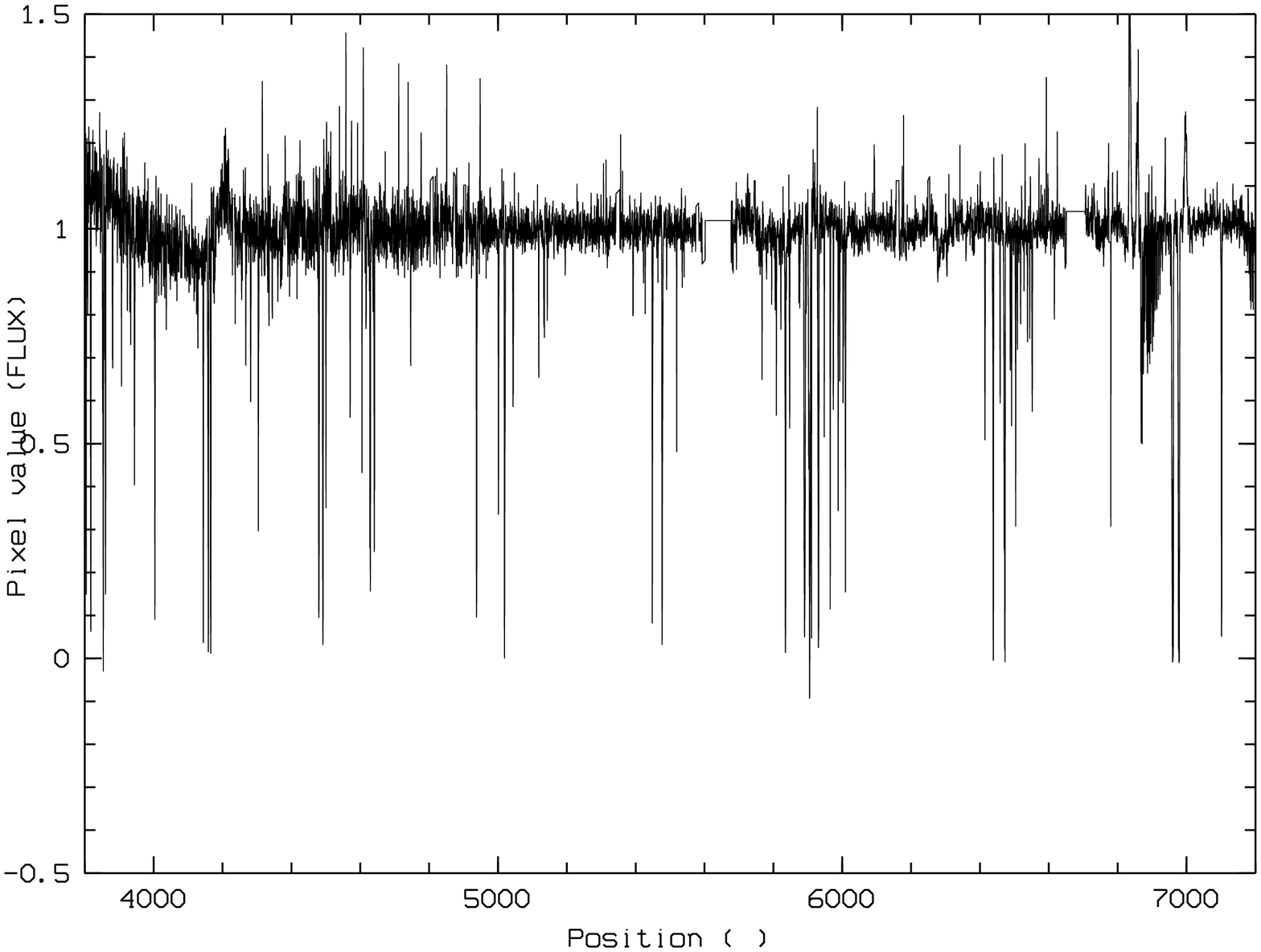}
}

\caption{The spectra of GRB050730, GRB050922C, GRB060418 observed with UVES@VLT.}
\label{ove}
\end{figure}

The high resolution spectrum of GRB050730, GRB050922C and GRB060418
are shown in Fig.\ref{ove}.  The resolution of all spectra is R 40,000
(7.5 km/s in the observer frame). Data sets were reduced using UVES
pipeline for MIDAS.  All afterglows are clearly detected in the range
3300 - 10000 \AA .

\section{Analysis}

For each absorption system several lines spread over the entire
spectral range covered by UVES observations were fitted. The strongest
absorption lines are observed in the host galaxy systems.  Several
components are identified (Fig. \ref{CIVSiIV}).  The lines were
analysed using the line fitting program FITLYMAN, part of the MIDAS
data reduction software package. FITLYMAN allows for the simultaneous
fitting of multiple absorption systems.

Both high and low ionization lines are detected, but the ratio of
their column densities is different for each component. This means
that the circumburst gas features have several ionization states
(Fig. \ref{CIIetc}).  As observed in previous high resolution
spectroscopy of GRB020813 and GRB021004 the circumburst environment
showed evidence of a clumpy environment, consisting of multiple shells
\cite{fiore05}.

\vspace{.5cm}

\begin{figure}[h!]
 \includegraphics[height=.37\textheight, angle=-90]{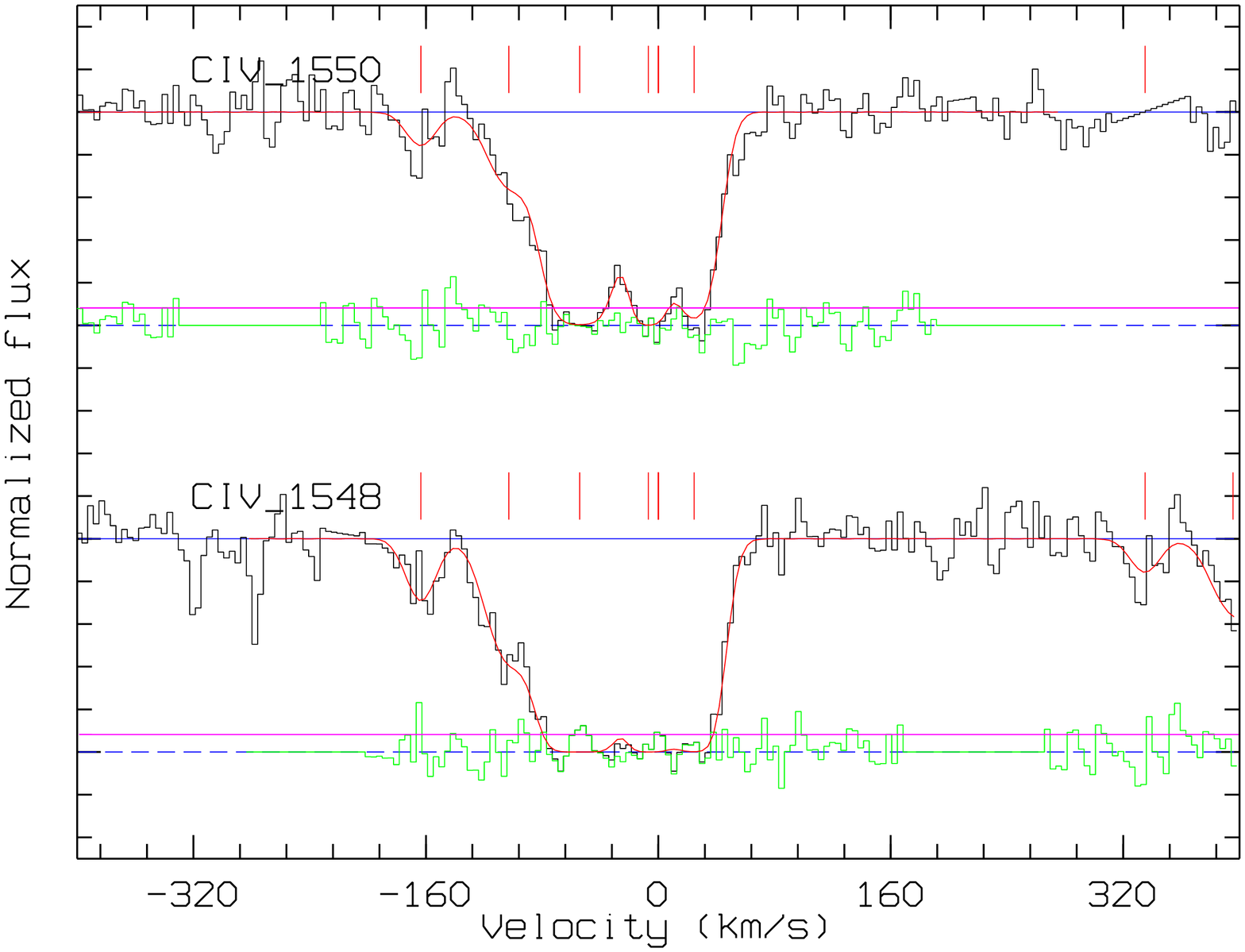}
 \includegraphics[height=.37\textheight, angle=-90]{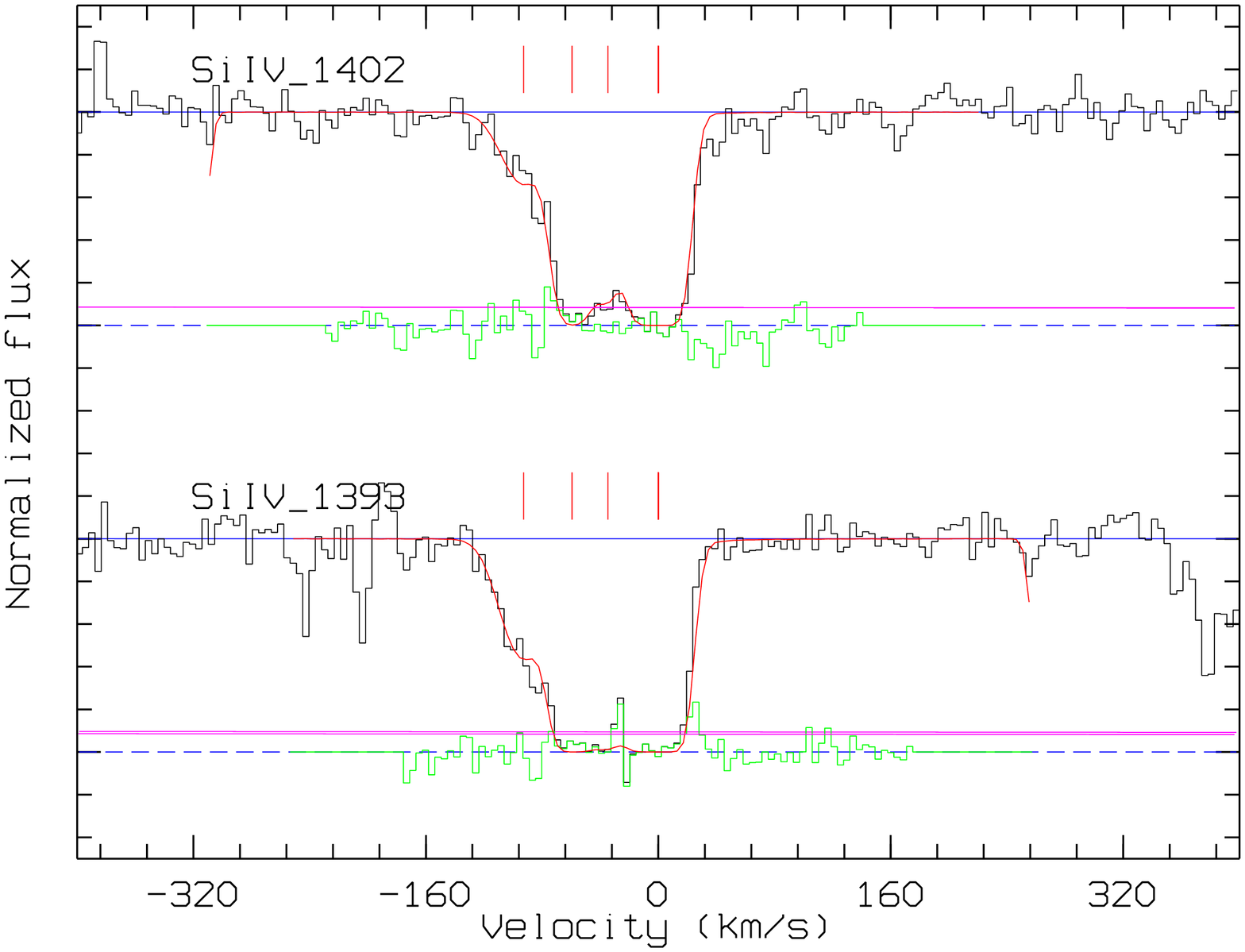}
  \caption{The CIV 1548 and 1550 \AA ~(left) and SiIV 1393 and 1402 \AA ~(right) doublets in
GRB050730.}
\label{CIVSiIV}
\end{figure}

\section{Fine Structure Lines}
Fine structure lines for CII, OI, FeII and SiII have been identified
in all the GRBs. Such lines convey information on the temperature
and electron density of the absorbing medium, provided that they are
excited by collisional processes (\cite{bahcall68}).

\begin{figure}[h!]
  \includegraphics[height=.37\textheight, angle=-90]{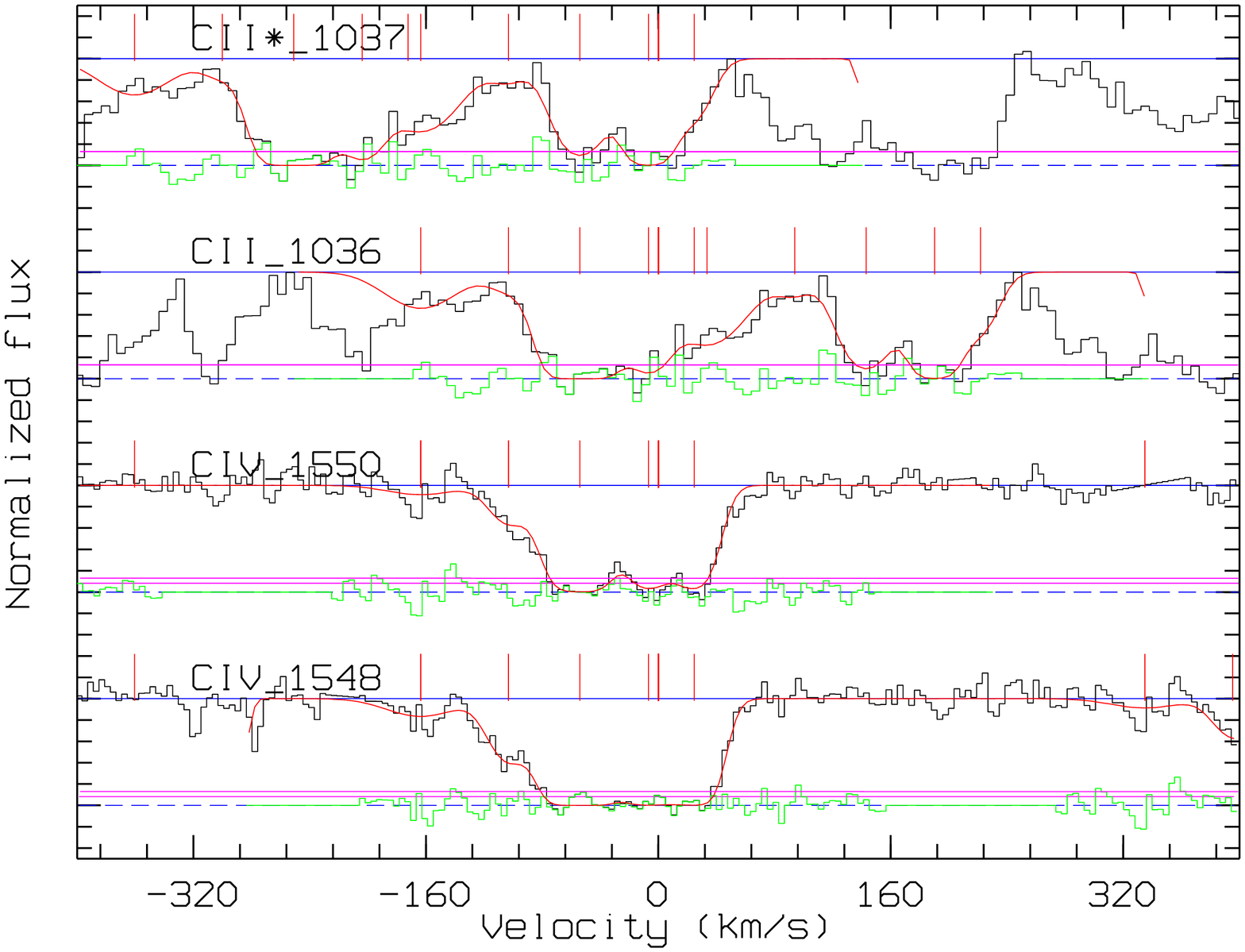}
  \includegraphics[height=.37\textheight, angle=-90]{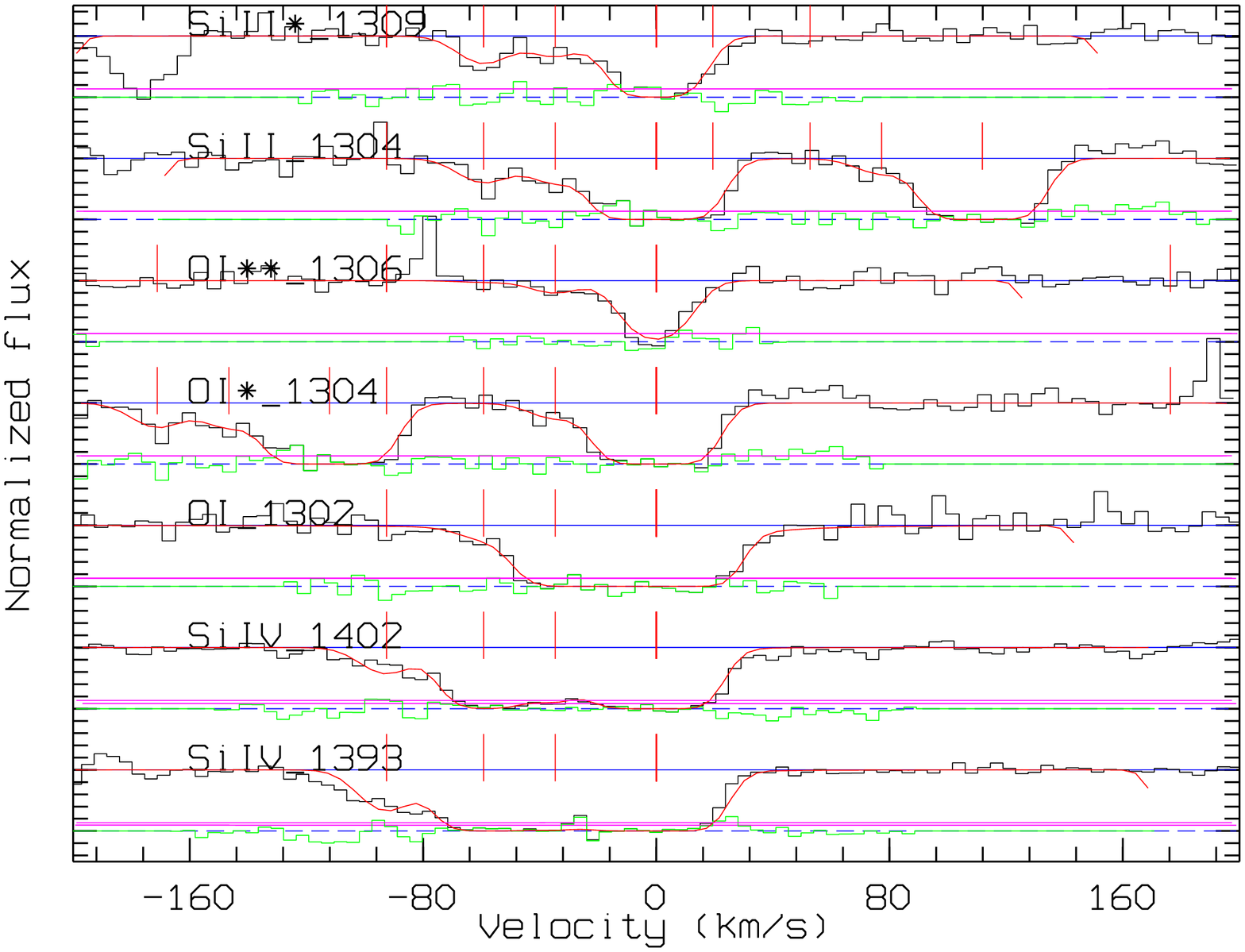}
  \caption{The low ionization lines CII, OI and SiII, together with
  their fine structure excited levels in GRB050730. Low ionization
  states and fine structure levels do not appear in all
  components. High ionization lines CIV and SiIV have been taken as
  reference lines in the fitting procedure.}
\label{CIIetc}
\end{figure}

To constrain these parameters we need to estimate the fine structure
column densities for two different ions and compare them. For GRB050730,
two out of five components show fine structure lines
(Fig. \ref{CIIetc}). Reliable values for temperature and electron density are T a
few $10^3$ K and $n > 10^4$ cm$^{-3}$ (second component; the
components are numbered according to decreasing z) and $n \sim 10
\div 100$ cm$^{-3}$ (third component). The other components do not
show fine structure features: this is an indication that they refer to a
clumpy environment.
\vspace{1cm}

\begin{figure}
  \includegraphics[height=.55\textheight, angle=-90]{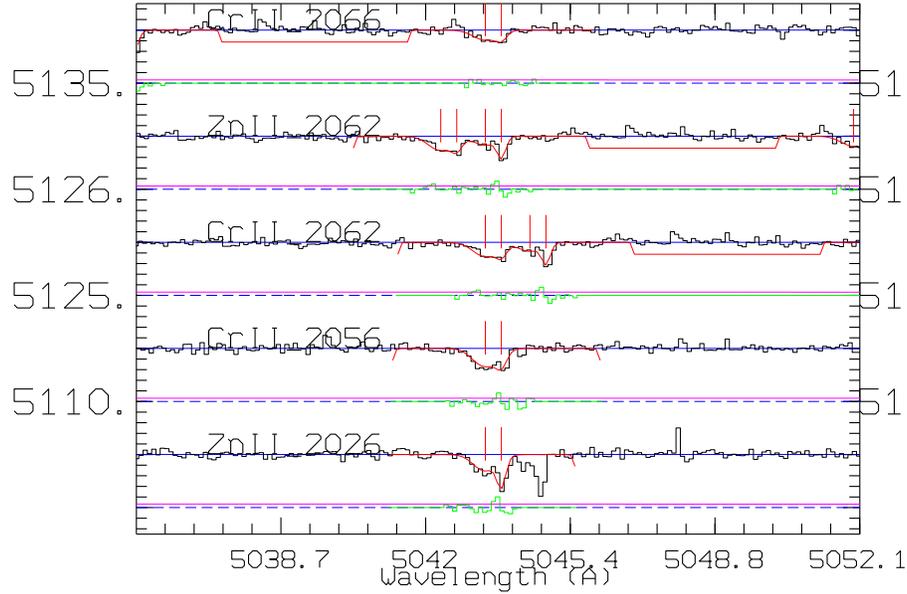}
  \caption{The CrII and ZnII absorption features in GRB060418}
\label{met}
\end{figure}

\section{Metallicity}

Metallicity in GRBs can be measured comparing the column densities of
heavy elements to that obtained for hydrogen by fitting the
Ly$-\alpha$, $\beta$ and $\gamma$ profiles. Both for GRB050730 and
GRB050922C, we find metallicities between $10^{-3}$ and $10^{-2}$ with respect
to the solar values.

Since metals tend to form dust, that then does not contribute to the
absorption lines, this result is affected by some
uncertainties. In GRB060418 we identify CrII and ZnII lines (Fig. \ref{met}). 
Such elements tend to stay in the gas state, minimizing the
uncertainty when estimating the metallicity. No H features are
present in this GRB spectrum, so we derive the $N_{H}$ column from the
X-ray data, leading to: Z(Cr) $= -1.8 \pm 0.3$ and Z(Zn) $= -1.3 \pm0.2$, a
bit higher than for the other two GRBs, but still below the solar
values.

Even if Cr and Zn features are not observed in the spectra, a correlation linking
FeII and ZnII abundances can be used \cite{Savaglio06}:

$$\log N_{ZnII} = (1.47 \pm 0.11) \log N_{FeII} + (-9.4 \pm 1.7).$$  

For GRB050730 we find Z(Zn) $= -1.5 \pm0.4$, while for GRB050922c we
obtain Z(Zn) $= -0.2 \pm 0.2$.  We show in Fig. 5 how the GRBs
analyzed in this paper place in the plot from \cite{Savaglio06} Such a
plot shows that GRB host galaxies (filled circles) have a higher
metallicity than galaxies lying along the line of sight of quasar
(open squares). Since the light coming from the quasars tends to
preferentially probe the halos of the intervening galaxies for a cross
section effect, while the GRB afterglows probe the innermost regions
of the hosts. Fig. 5 shows that GRB050730 has a value of Z in
agreement with the redshift-metallicity relation, while that of
GRB060418 lies beyond the correlation. This could be due to the fact
that in the latter case the H column has been computed from the X-ray
data, since no optical features were available in the spectrum.

\begin{figure}
  \includegraphics[height=.55\textheight, angle=0]{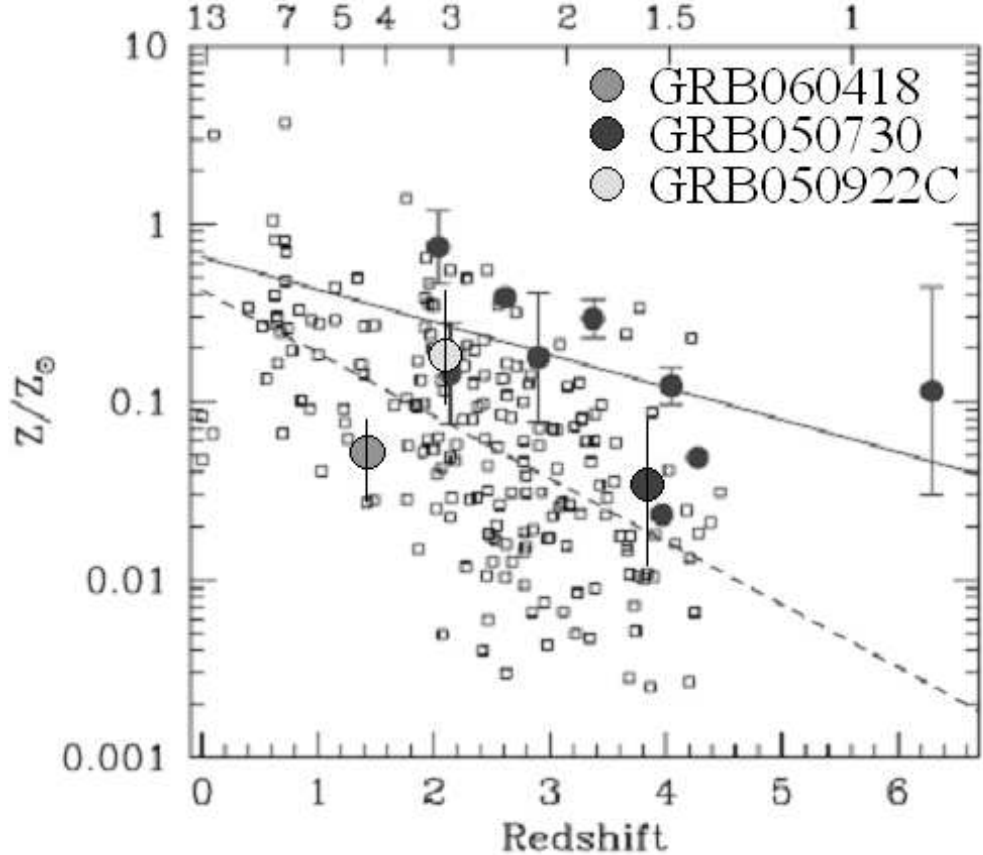}
  \caption{The metallicity redshift relation for galaxies lying along
  the line of sight of quasars (open squares), for GRB host galaxies
  (filled circles) and for GRB050730, GRB050922C and GRB060418 (filled
  triangles). The metallicity for GRB hosts has been computed using
  the Zn spectral features when present, or the FeII/Zn correlation
  otherwise. The plot is taken from from \cite{Savaglio06}. }
\end{figure}

\section{The [C/Fe] Ratio}

We can also calculate the ratio [C/Fe], representative of the
enrichment of the $\alpha$ elements relative to iron. For GRB050730 we
measure a mean [C/Fe]=0.08$\pm$0.24, consistent with the value
predicted by the models of \cite{Pipino06} for a
galaxy younger than 1 Gyr subject to a burst of star-formation.  In
such a case \cite{Pipino06} also predict a low [Fe/H]
value, close to what we find. Again, we warn that mean abundance
estimates may be affected by large systematic uncertainties.

\begin{figure}[h!]
\vbox{ 
 \includegraphics[height=.55\textheight, angle=-90]{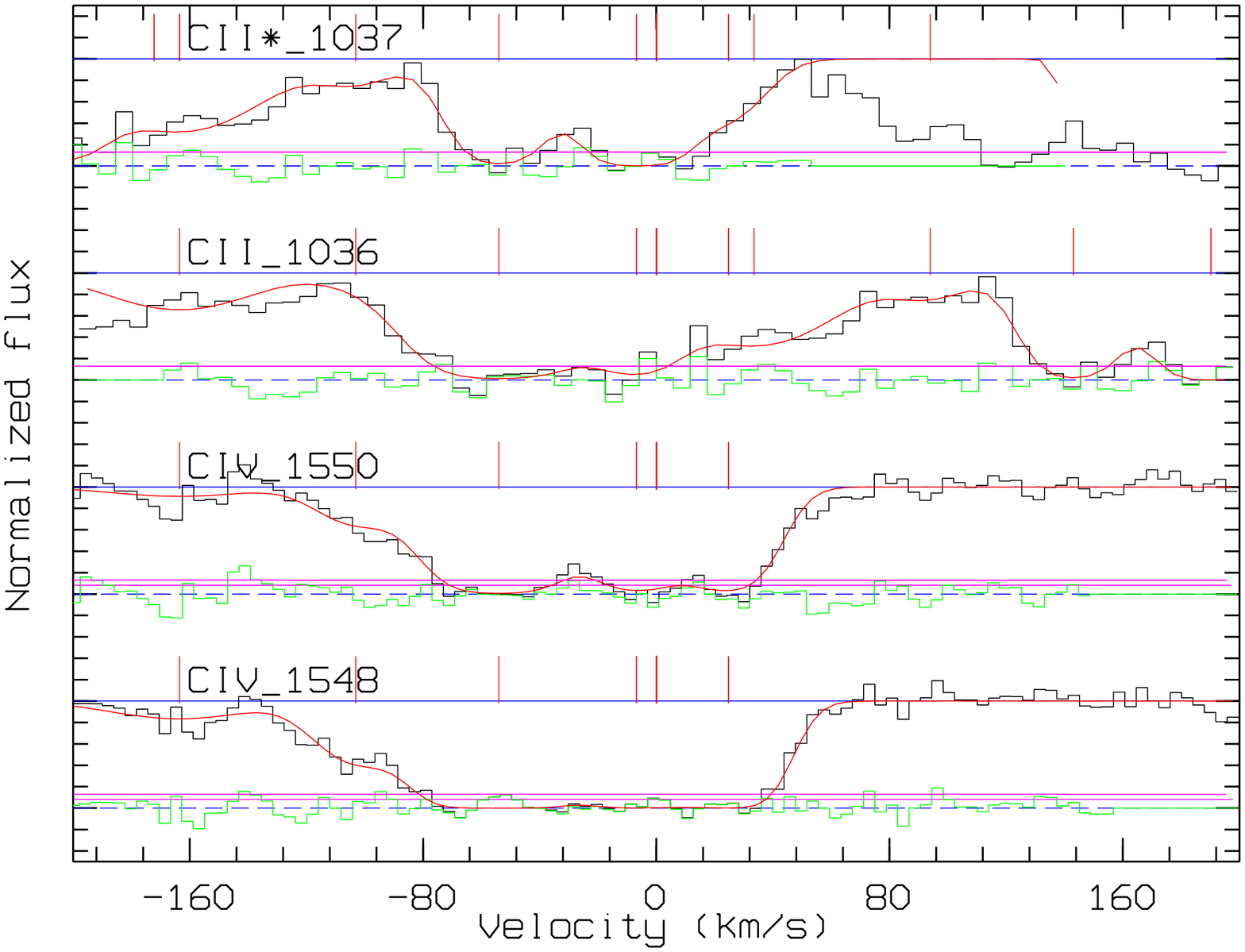}

 \includegraphics[height=.55\textheight, angle=-90]{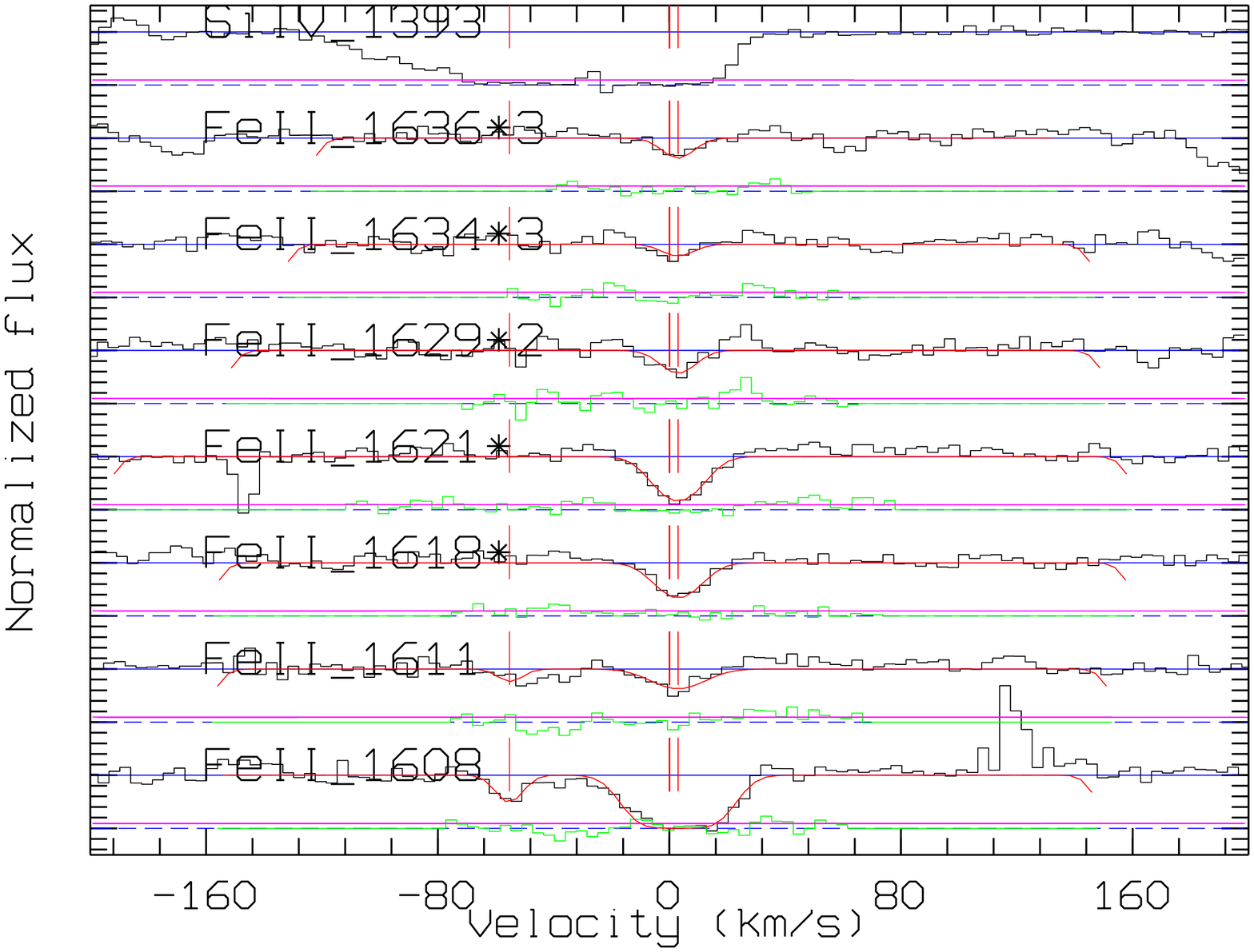}

}
  \caption{The CIV and FeII absorption features in GRB050730. Note that the CII is
roughly constant in component 2 and 3, while the FeII is almost absent in component 3.}
\label{CIVSiIV}
\end{figure}

A safer approach is to estimate the [C/Fe] of each component of the
main absorption system. For GRB050730, we observe FeII columns only on
component 2 and 3 (the same in which fine structure lines are
present). We find that [C/Fe] of component 2 is -0.15$\pm$0.13 while
that of component 3 is +0.53$\pm$0.23.  More specifically, while the
total carbon column density of component 2 and 3 are similar, the
total colum density of iron of component 3 is about four times less
than that of component 2 (see Fig. 6). A similar conclusion applies to silicon.
The total silicon column density of component 3 is also about 10 times
less than that of component 2.  Interestingly, \cite{Perna03} found
that silicates tend to be destroyed more efficiently than graphite if
a dusty medium is exposed to the intense GRB radiation field. This
would leave more iron free in the gas phase in clouds closer to the
GRB site than in farther clouds, possibly indicating that component 2
is closer to the GRB site than component 3.

\section{Conclusions}
The absorption spectra of GRB afterglows are extremely complex,
featuring several systems at different redshifts.  Both high and low
ionization lines are observed in the circumburst environment, but
their relative abundances vary from component to component, indicating
a clumpy environment consisting of multiple shells.  Fine structure
lines give information on the temperature and electron density of the
absorbing medium, provided that they are excited by collisional
effects. Different components have different densities, suggesting a
variable density profile. Reliable values for temperature are T a few
$10^3$ K, while density varies from component to component. For
GRB050730 we find $n > 10^4$ cm$^{-3}$ in component 2 and $n \sim 10
\div 100$ cm$^{-3}$ in component 3, respectively. Metallicity can be derived
from the metal column densities relative to that of Hydrogen. Values
between $10^{-3}$ and $10^{-2}$ with respect to solar are
found. However, since metals tend to form dust, that then does not
contribute to the absorption lines, this result is affected by some
uncertainties. Cr and Zn are the best indicators, since they do not
form dust. Even if these indicators are not observed in the spectra,
we can compute the column of ZnII using a correlation law that links
such a ion with the FeII. Metallicity values above $10^{-2}$ with
respect to the solar ones have been found with this method. Finally,
the [C/Fe] ratio can give us informations about the galaxy
evolution. For GRB050730, whose redshift is $z = 3.967$, we find a
value in agreement with that expected for a galaxy younger than 1 Gyr
subject to a burst of star-formation. Again, dust depletion may still
play an important role. The [C/Fe] ratio evaluated component by
component in GRB050730 shows that FeII is less abundant in component 3
than in component 2. This indicates that the gas in the latter is
closer to the GRB than that in the former, since iron dust is more
efficiently destroyed than graphite.

More details can be found in \cite{delia06}.

\end{document}